\documentclass[twocolumn,prl,amssymb, aps,superscriptaddress,
  showpacs,preprintnumbers, amsmath,floatfix]{revtex4}

\usepackage{graphicx}
\usepackage{color}

\newcommand{\be}{\begin{equation}}
\newcommand{\ee}{  \end{equation}}
\newcommand{\ba}{\begin{eqnarray}}
\newcommand{\ea}{  \end{eqnarray}}

\begin{document}

\title{Neutron Resonance Widths and the Porter-Thomas Distribution}

\author{Alexander \surname{Volya}}
\email{avolya@fsu.edu}
\affiliation{Department of Physics, Florida State University, Tallahassee, Florida 32306-4350, USA}

\author{Hans A. \surname{Weidenm{\"u}ller}}
\affiliation{Max-Planck-Institut f\"ur Kernphysik, Saupfercheckweg 1, D-69117 Heidelberg, Germany}

\author{Vladimir \surname{Zelevinsky}}
\affiliation{NSCL and Department of Physics and Astronomy, Michigan State University, East Lansing, Michigan 48824-1321, USA}


\begin{abstract} Experimental evidence has recently put the validity
of the Porter-Thomas distribution (PTD) for partial neutron widths
into question. We identify two terms in the effective Hamiltonian that
violate orthogonal invariance (the basis for the PTD). Both are due to
the coupling to the decay channels. We show that realistic estimates
for the coupling to the neutron channel and for non-statistical gamma
decays yield significant modifications of the PTD.
\end{abstract}

\maketitle

{\it Introduction.} Recent experimental results on the distribution of
neutron resonance widths have cast serious doubt on the validity of
random-matrix theory (RMT) in nuclei. RMT predicts that the reduced
neutron widths follow a Porter-Thomas distribution (PTD)~\cite{Por56}
(a $\chi^2$ distribution with a single degree of freedom). That
prediction  assumes non-overlapping resonances
with a single open channel (the neutron channel). Deviations from the
PTD for several open channels are analyzed in Refs.~\cite{Shc12,
  Fyo15}. Agreement with the PTD could be excluded with very high
probability both for the Pt isotopes~\cite{Koe10} and for the Nuclear
Data Ensemble~\cite{Koe11}. Numerous theoretical attempts~\cite{Wei10,
  Cel11, Vol11} to account for the results of Refs.~\cite{Koe10,
  Koe11} have not definitively resolved the issue. The validity of RMT is of
central importance for the statistical theory of nuclear
reactions~\cite{Mit10} that is widely
used in nuclear cross-section calculations.

In the present paper we address two dynamical effects that modify the
PTD and that apparently have not been taken into account so far in the
theoretical literature or in the analysis of neutron resonance
data. The two effects are the Thomas-Ehrman shift known from the study
of light nuclei~\cite{Tho52}, and non-statistical effects in the 
gamma decay of the neutron resonances~\cite{Koe13}.  We show that both
may cause significant deviations of the distribution of neutron
resonance widths from the PTD.

The PTD follows from the orthogonal invariance of the Gaussian
Orthogonal Ensemble of random matrices (the GOE). The GOE distribution
function is proportional to
\ba
{\rm d} {\cal O} \exp \bigg\{ - \frac{N}{\lambda^2}
\sum_\rho E^2_\rho \bigg\}\prod_{\mu < \nu}^N |E_\mu - E_\nu| \prod_{\sigma}^N
{\rm d} E_\sigma \ .
\label{1}
\ea
Here $N$ with $N \to \infty$ is the dimension of the GOE matrices. The
parameter $\lambda$ defines the width $4 \lambda$ of the GOE
spectrum. The mean level spacing at the center of the GOE spectrum is
$d = \pi \lambda / N$. The $E_\mu$ are the GOE eigenvalues, and ${\rm
  d} {\cal O}$ is the Haar measure of the orthogonal group in $N$
dimensions.  
{It encompasses the GOE eigenfunctions.}
Factorization of
the distribution implies that eigenvalues and eigenfunctions are
statistically independent. 
For $N \to \infty$, the projections of the
  eigenfunctions onto an arbitrary vector in Hilbert space possess a
Gaussian distribution, and the reduced widths, therefore, have a
PTD. The reported disagreement of the distribution of reduced neutron
widths with the PTD directly challenges the postulated orthogonal
invariance of the GOE. Conversely, dynamical effects that violate
orthogonal invariance will cause deviations from the PTD. We show that
both, the Thomas-Ehrman shift and non-statistical gamma decays, have
that property and, thus, qualify as causes of the observed
disagreement.

{\it Violation of Orthogonal Invariance.} We consider $s$-wave neutron
scattering on a spin zero target nucleus. We take account of the
single open neutron channel and of the large number of gamma decay
channels. The effective Hamiltonian is~\cite{Mit10}
\ba
H^{\rm eff}_{\mu \nu} &=& H^{\rm GOE}_{\mu \nu} + F_{\mu \nu}(E) - i \pi W_\mu(E)
W_\nu(E) \nonumber \\
&& \qquad - i \pi \sum_\gamma W_{\mu}^{(\gamma)} W_{\nu}^{(\gamma)} \ .
\label{2}
\ea
Here $E$ is the neutron energy. We have replaced the actual
Hamiltonian by the GOE Hamiltonian $H^{\rm GOE}$ as defined in
Eq.~(\ref{1}). The real matrix elements $W_\mu(E)$ with $\mu = 1, 2,
\ldots, N$ couple the $s$-wave neutron channel to the space of $N$
resonance states and carry the same energy dependence $E^{1/4}$ as do
the neutron partial width amplitudes. The shift matrix $F$ accounts
for that energy dependence. It is the analog of the Thomas-Ehrman
shift, with elements
\be
F_{\mu \nu}(E) = {\rm Pv} \int_0^\infty {\rm d} E' \ \frac{W_\mu(E')
W_\nu(E')}{E - E'}
\label{3}
\ee
where ${\rm Pv}$ indicates the principal-value integral. At energies
far above threshold the matrix $F$ is often neglected because then
contributions to the integral from energies $E' < E$ and $E' > E$ tend
to cancel. Such cancellation cannot occur at neutron threshold $E = 0$
and $F_{\mu \nu}(E)$ may, thus, not be negligible. We neglect
contributions similar to $F$ from closed channels and take $F(E)$ as a
paradigmatic example. The matrix elements  $W_{\mu}^{(\gamma)}$ play the same role for the gamma channels as do the $W_\mu$
for the neutron channel except  
for a different energy dependence of the
$W_{\mu}^{(\gamma)},$ resulting in a negligible contribution
to the Thomas-Ehrman shift.

None of the terms added to $H^{\rm GOE}$ in Eq.~(\ref{2}) is invariant
under orthogonal transformations. 
Addressing the regime of isolated resonances
we confine our attention to the matrix
$F$ and to the coupling to the gamma decay channels.
Thus, the elements of the
width matrix $W_\mu W_\nu$ only serve to define the neutron decay
widths and are otherwise negligible. Beyond that regime, the width
matrix does cause deviations from the PTD~\cite{Cel11, Vol11}.

Let us first disregard gamma decay channels and concentrate on the role of $F$. Writing
$W_\mu(E) = {\cal W}_\mu E^{1/4}$ we observe that the matrix
${\cal W}_\mu {\cal W}_\nu$ has a single nonzero eigenvalue $\sum_\mu
{\cal W}_\mu^2$. The associated eigenvector defines the superradiant
state~\cite{Sok88, Sok92} labeled $\mu = 1$. The transformation to the
eigenvector basis leaves the ensemble of GOE matrices unchanged and
yields 
\be 
H^{\rm eff}_{\mu \nu} \approx H^{\rm GOE}_{\mu \nu} + \delta_{1 \mu}
\delta_{1 \nu} \left [ F - i \pi \sum_\rho W^2_\rho(E) \right ] \label{4}
\ee
where  $F=\sum_\mu
F_{\mu \mu}.$ 

%
The transformation does not diagonalize the matrix $F$ exactly because
the integral defining $F$ receives contributions also from higher
energies where the approximation $W_\mu(E) \approx {\cal W}_\mu
E^{1/4}$ does not apply. Near neutron threshold such contributions are
relatively small, however, and Eq.~(\ref{4}) should be a good
approximation.

We compare $F$ (Eq.~(\ref{4})) with the diagonal element $H^{\rm
  GOE}_{1 1}$. Sufficiently far above neutron threshold (where the
$s$-wave penetration factor is $\approx 1$) the matrix elements
$W_\mu$ obey~\cite{Mit10} $\sum_\mu W^2_\mu = N d x / \pi^2 $ where
$x$ is related to the average of the $s$-wave scattering function via
$\overline{S} = (1 - x) / (1 + x)$ and is, thus, of order unity. Hence
$\sum_\mu W^2_\mu \approx d N / \pi^2 = \lambda / \pi$. For $N \gg 1$
that is much larger than the root-mean-square value $\lambda \sqrt{2 /
  N}$ of the diagonal element $H^{\rm GOE}_{1 1}$. For $E \to 0$ the
$s$-wave penetration factor reduces $\sum_\mu W^2_\mu$. The reduction
does not affect the principal-value integral, however, which actually
attains its maximum value at $E = 0$. Thus, we expect $F \approx
\lambda / \pi$. The effect of $F$ can be amplified beyond our estimate by
a single-particle resonance near neutron threshold. Such is the case
in the Pt isotopes for the $4 s$-state of the shell
model~\cite{Wei10}.

Next, we turn to the gamma channels in Eq.~(\ref{2}).
According to the statistical model, the matrix elements $W_{\mu}^{(\gamma)}$ are Gaussian-distributed random variables. The total gamma
widths of the neutron resonances are, therefore, expected to have a
$\chi^2$ distribution with a large number of degrees of freedom, and
the partial widths $\Gamma_{\mu}^{(\gamma)}$ of the neutron resonances in a given
target nucleus are expected to have all nearly the same value. The
effective Hamiltonian is, thus, expected to be approximately given by
$H^{\rm eff}_{\mu \nu} = H^{\rm GOE}_{\mu \nu} - i \pi W_\mu W_\nu -
\delta_{\mu \nu} i \overline{\Gamma} / 2$, with $\overline{\Gamma}$
independent of $\mu$. The term $\delta_{\mu \nu} i \overline{\Gamma} / 2$
is obviously orthogonally invariant. While the contribution of each
gamma channel to $\overline{\Gamma}$ is small the number of such channels
is large resulting in a value of $\overline{\Gamma}$ that dominates the
total neutron resonance widths near neutron threshold.

A recent analysis~\cite{Koe13} of the distribution of total gamma
decay widths of neutron resonances in $^{66}$Mo contradicts the
expectation that these all have the same value. For $s$-wave
resonances and positive parity states, the distribution is shown in
the lower part of Fig.~6 of Ref.~\cite{Koe13}. The distribution is
much wider than predicted by the GOE. The result confirms earlier
data~\cite{She07} comprising a much smaller number of resonances. It
seems that at present, the cause for the deviation is not
understood. It is not clear whether it is due to a specific property
of $^{66}$Mo or whether it is likely to occur in other nuclei as
well. We opt for the second possibility. We assume that the total
gamma decay widths generically possess large fluctuations, in
contradiction to the statistical model. We explore the consequences of
that hypothesis for the distribution of neutron decay widths.

Typical values of total gamma decay widths for $s$-wave neutron
resonances are of order $100$ meV, both in medium-weight~\cite{Koe13}
and in heavy~\cite{Gar64} nuclei. The distributions for the total
gamma decay widths shown in Ref.~\cite{Koe13} start roughly at
$\Gamma_0 = 100$ meV and fall off with a half width $\sigma$ of
roughly $300$ meV. To transcribe these figures into the effective
Hamiltonian of Eq.~(\ref{2}) we use that in medium-weight and heavy
nuclei typical average resonance spacings $d = \pi \lambda / N$ are of
order $10$ eV. Then $\Gamma_0 \approx \pi \lambda / (100 N)$ and
$\sigma \approx \pi \lambda / (30 N)$. The orthogonal invariance of
the coupling to the gamma decay channels is broken by the spread
$\sigma$. We compare $\sigma$ with the Thomas-Ehrman shift function $F
\approx \lambda / \pi$. We have $(1 / N) {\rm Tr} \ \sigma \approx \pi
\lambda / (30 N)$ while $(1 / N) {\rm Tr} \ (F \delta_{\mu 1}
\delta_{\nu 1}) = F / N = \lambda / (\pi N)$. While somewhat smaller,
gamma decay breaks orthogonal invariance roughly as strongly as does
the Thomas-Ehrman shift.

In summary we have identified two effects that break the orthogonal
invariance of the GOE, the Thomas-Ehrman shift and the non-statistical
distribution of gamma decay widths. In the remainder of the paper we
investigate the consequences of both effects for the distribution of
neutron resonance widths. We do so in the framework of a schematic
model. We write the effective Hamiltonian as
\be
H^{\rm eff} = H^{\rm GOE}_{\mu \nu} + \delta_{\mu 1} \delta_{\nu 1} Z .
\label{5}
\ee
%
We have suppressed the constant gamma decay width $\overline{\Gamma}$
because it only causes a uniform shift of all eigenvalues \cite{Shc12} and does not
affect the eigenfunctions. The constant
\be
Z = F - (i/2) \delta \Gamma-i \pi \sum_\rho W^2_\rho
\label{6}
\ee
violates orthogonal invariance and includes the Thomas-Ehrman
shift; the effect of non-statistical gamma
decays where we schematically  represent the spread of gamma decay widths by a single
term; and the term $- i \pi \sum_\rho W^2_\rho$ which is a reminder that we use
a basis where the state $| 1 \rangle$ is the superradiant state.

{\it Average Level Density.} It is easy to see that for $N \to \infty$
the presence of $Z$ in the effective Hamiltonian has a negligible
influence on the average level density. The level density is defined
as $\rho(E) = - (1 / \pi) \Im (E^+ - H^{\rm eff})^{- 1}$. Here $(E^+ -
H^{\rm eff})^{- 1}$ is the retarded Green function. This expression for
$\rho$ is physically meaningful only if $H^{\rm eff}$ is Hermitian,
i.e., if $Z$ is real. We first consider that case and put $Z = F$.
The term $F$ in $H^{\rm eff}$ may be considered as causing a doorway
state.  Therefore, we treat the first line and the first column of
$H^{\rm eff}$ differently from the rest. We define the orthogonal
projection operators $P = | 1 \rangle \langle 1 |$ and $Q = 1 -
P$. With $\mu, \nu \geq 2$ we write $P H^{\rm eff} P = H^{\rm GOE}_{1
  1} + Z = E_0$, $( P H^{\rm eff} Q )_{1 \mu} = H^{\rm GOE}_{1 \mu} =
V_\mu$, $( Q H^{\rm eff} Q )_{\nu 1} = H^{\rm GOE}_{\nu 1} = V_\nu$,
and $( Q H^{\rm eff} Q )_{\nu \mu}$. The elements of the matrix $Q
H^{\rm eff} Q$ are Gaussian-distributed random variables. Moreover,
the probability distribution of $Q H^{\rm eff} Q$ is invariant under
orthogonal transformations in $Q$-space. Therefore, the matrices $Q
H^{\rm eff} Q$ form a GOE of dimension $N - 1$. We denote that
ensemble by $\tilde{H}^{\rm GOE}$ and, suppressing the term $\sum_\rho
W^2_\rho$, write the total Hamiltonian in matrix form,
\be
H^{\rm eff} = \left( \begin{matrix} E_0 & V_\mu \cr
                 V_\nu & \tilde{H}^{\rm GOE}_{\nu \mu} \cr \end{matrix}
            \right) \ .
\label{7}
\ee
The right-hand side of Eq.~(\ref{7}) is identical in form with the
standard model for a doorway-state (see, for instance,
Ref.~\cite{Dep07}).  The spreading width of the doorway state is
$\Gamma^{\downarrow} = 2 \pi (1/N) \sum_\mu V^2_\mu / d$. In the
present case we have $\sum_\mu V^2_\mu = \sum_\mu H^2_{1\mu}$. We note
that for $N \gg 1$ the sum is self-averaging and given by $\lambda^2$
for every member of the ensemble~(\ref{7}). The resulting value of the
spreading width $\Gamma^{\downarrow} = 2 \lambda$ is comparable with
the total width of the GOE spectrum, and the doorway state is
completely smeared over that spectrum. While in the standard
doorway-state model~\cite{Dep07}, the magnitude of the coupling matrix
elements $V_\mu$ is of order $d$, here their root-mean square values
$d \sqrt{N}$ are fixed by the underlying GOE and very large compared
to $d$. Therefore, the doorway state does not cause a local
enhancement of the level density. We note that the model of
Eq.~(\ref{7}) is physically meaningful only for $|Z| \leq 2
\lambda$. For $|Z| \gg 2 \lambda$ there exists a distinct state
outside the GOE spectrum that carries (almost) all the coupling to the
neutron channel. That case does not seem to model the scattering of
slow neutrons in a meaningful way.

As a corollary we mention that for imaginary $Z$ and $|Z| \approx 2
\lambda$ we deal with a superradiant state that causes a pole of the
Green function in the complex energy plane. The distance of that pole
from the real axis is comparable with the width of the GOE spectrum.
Therefore, the superradiant state conveys only a small part of its
large neutron width to the remaining GOE eigenstates. 
It does not seem meaningful to
consider imaginary values of $Z$ in excess of $ 2 \lambda$.

{\it Numerical Results.} With $\phi_\mu$ the normalized eigenfunctions
of the effective Hamiltonian~(\ref{5}) with  $\sum_\rho W^2_\rho
  \rightarrow 0$, the partial neutron decay widths are proportional
to $N |\langle 1 | \phi_\mu \rangle|^2$. The factor $N$ is introduced
so that the average width equals unity. We have
calculated the effect of $Z$ on the distribution of partial
  widths $x=N |\langle 1 | \phi_\mu \rangle|^2$ perturbatively, both
for $|Z| \ll d$ and for $|Z| \gg \lambda$. The results show that $Z$
does influence the PTD. Both limits are unrealistic, however, and
serve only as a check for the numerical work. Therefore, we do not
give our analytical results here.

\begin{figure}[h]
\includegraphics[width=0.7\linewidth]{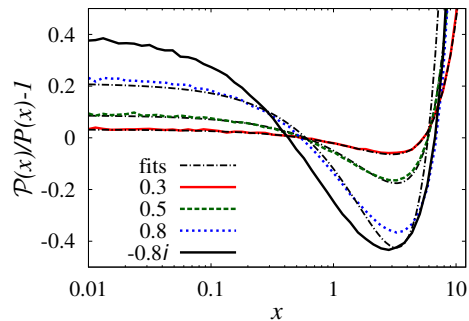}
\caption{(Color online) Relative difference $[{\cal P}(x) / P(x)] - 1$ as a function
  of $x$ for several values of $\kappa$ as indicated in the
  figure and for $N = 1000$. The dash-dot lines are fits using
  expression~(\ref{9}).
\label{fig:gamma}
}
\end{figure}

We first present results for real $Z$. We use the dimensionless
parameter $\kappa = Z / \lambda$ and consider values that are
physically realistic but lie outside the range of validity of the
perturbative approach.  As $Z$ increases the PTD $P(x) = (1 / (2
\sqrt{2 \pi x})) \exp \{ - x / 2 \}$ 
is deformed. The
resulting probability distribution ${\cal P}(x)$ of the partial widths
is shown in a plot of $[{\cal P}(x) / P(x)] - 1$ versus $x$ for
several values of $\kappa$ in Fig.~\ref{fig:gamma}. The term $Z | 1
\rangle \langle 1 |$ in $H^{\rm eff}$ leads to a segregation of
states. The states in one group become broader and those in the other
group become more narrow. Therefore, the modified distribution has a
longer tail and is more strongly peaked at $x = 0$ than the PTD, while
in the middle around $x = 1$ the distribution is suppressed. In the
limit of very large $\kappa$ one state becomes collective (i.e.,
carries almost all the decay strength). The distribution of widths of
the remaining states returns to the PTD but with a much reduced
average width. That is consistent with our perturbative results for
large $Z$.

We have fitted these curves using the parametrization
\be
{\cal P}(x) = [1 + A(1 - x) + B(x^2 -6 x + 3)] P(x)
\label{8}
\ee
suggested by the perturbative result for small $Z$. The distribution
${\cal P}(x)$ in Eq.~(\ref{8}) is normalized to unity for all values
of $A$ and $B$. Terms linear (quadratic) in $x$ embody a change in
average width (a quenching of the PTD, respectively). For each of the
curves representing the data we also show a fit using Eq.~(\ref{8}) as
a dash-dotted line. In the interval $0.1 \leq \kappa \leq 1$ that is
most relevant for applications, the deviations from the PTD are
described perfectly by expression~(\ref{8}).

\begin{figure}[h]
\includegraphics[width=0.8\linewidth]{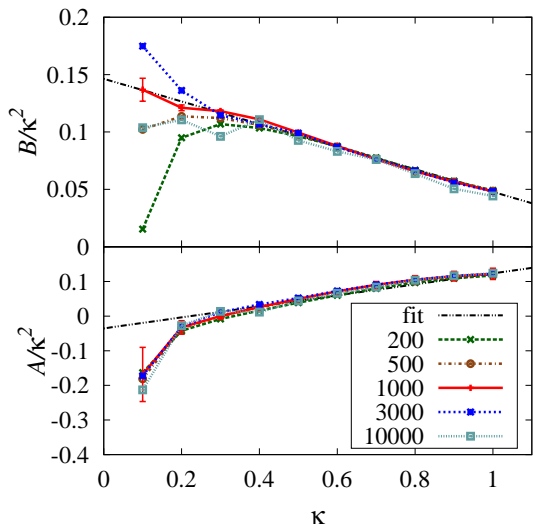}
\caption{(Color online) Scaled fit coefficients $A/\kappa^2$ and $B/\kappa^2$ with
  $A$ and $B$ defined in Eq.~(\ref{9}) are plotted as functions of
  $\kappa$ for GOE ensembles of different dimension $N$ as indicated
  in the figure. For $N = 1000$ the error in the fit is shown by error
  bars. The errors for the other curves are similar. The
  double-dot-dash black line shows the linear fits~(\ref{9}) of
  $A/\kappa^2$ and $B/\kappa^2$ as functions of $\kappa.$
\label{fig:abplot}
}
\end{figure}

The scaled fit coefficients $A / \kappa^2$ and $B / \kappa^2$ are
shown as functions of $\kappa$ in Fig.~\ref{fig:abplot} for different
matrix dimension $N$. For $\kappa < 0.4$ we encounter numerical
instabilities. However, in that regime the changes of the PTD are
too small to be of practical interest. For $\kappa > 0.4$ the results
for different matrix dimension $N$ exhibit consistently a linear
dependence on $\kappa$ approximately given by
\ba
A / \kappa^2 &=& -0.035 \pm 0.010 + (0.16\pm 0.01) \kappa \ ,
\nonumber \\
B / \kappa^2 &=& 0.146\pm 0.002 + (0.099\pm 0.003) \kappa \ .
\label{9}
\ea

As a single quantitative measure of deviations from the PTD we use the
coefficient of L variation (also known as the Gini coefficient)
defined as
\be
\tau = \frac{1}{2\overline{x}} \, \int dx \int dx' \, |x-x'|
{\cal P}(x) \, {\cal P}(x').
\label{10}
\ee
The advantage of using $\tau$ over the traditional coefficient of
variation is that $\tau$ is nearly insensitive to the existence of a
small fraction of highly collective states that may comprise the
average value of $x$~\cite{Vol11}. The coefficient $\tau$ ranges
between zero and unity. For a $\chi^2$ distribution with $\nu = 1$ and
$\nu = 2$ we have $\tau = 2 / \pi$ and $\tau = 1 / 2,$ respectively;
$\tau$ decreases with increasing $\nu$. A value of $\tau > 2 / \pi
\approx 0.64$ corresponds to a distribution that is more strongly
peaked at small $x$ than the PTD and effectively has $\nu < 1$. For a
distribution of the form~(\ref{9}) we find $\tau = (2 / \pi) (1 - 2 A
- A^2 + 2 B - 2 AB - 3 B^2)$. For $\kappa=0.8,$ for
  example, Eq. (\ref{9}) yields $\tau=0.70$, in agreement with
  numerical tests. That value of $\tau$ has a $\chi^2$ distribution
  with $\nu = 0.72.$ 

The results for purely imaginary $Z$ are qualitatively similar to
those for real $Z$. Compared to the PTD, the distribution is increased
for small and large $x$ and is depressed for $x \approx 1$. That is
shown for  $Z/\lambda = - 0.8 i$ by the black solid line in
Fig.~\ref{fig:gamma}. For imaginary $Z$ we do not present a fit
because we have not succeeded in finding similarly good fit formulas
as in Eqs.~(\ref{9}) for real $Z$. 
{\it Conclusions.} The PTD follows from the orthogonal invariance of
the GOE. We have identified two causes for violation of that
invariance: The Thomas-Ehrman shift and non-statistical gamma decays.
We have shown that reasonable estimates for both cause significant
deviations of the distribution of neutron decay widths from the PTD.

Invariance breaking by the Thomas-Ehrman shift is due to the coupling
to the neutron channel. Such coupling is immanent in the theory and
does not invalidate the GOE. The shift is strongest at neutron
threshold and is expected to be particularly pronounced when the
$s$-wave strength function is maximal. In contradistinction, it may be
argued that the existence of non-statistical gamma decays represents a
genuine violation of GOE assumptions. It is conceivable that
such decays are due to transitions to low-lying states where
random-matrix theory does not apply. Hopefully time will tell.

In all cases studied, invariance breaking results in a depletion of
the probability distribution for the partial widths $x$ near $x = 1$,
compensated by an increase for small and large values of $x$. To
estimate the effect of such breaking in an individual nucleus, the
quantities $F$ and $\delta \Gamma$ in Eq.~(\ref{6}) must be
estimated. For $F$ that should be possible using the neutron strength
function. For $\delta \Gamma$ (a measure of the spread of total gamma
decay widths) the simultaneous analysis of the distribution of partial
neutron widths and of total gamma decay widths for the measured chain
of neutron resonances is required.

This material is based upon work supported by the U.S. Department of Energy Office of
Science under Award Number DE-SC0009883 and by the NSF grants PHY-1068217 and PHY-1404442.

\end{document}